\documentclass{iopconfser}
\usepackage[
    backend=biber,
    style=vancouver,
]{biblatex}
\usepackage[font=small,skip=2pt]{caption}
\usepackage{amsmath}

\usepackage{graphicx}
\usepackage{soul}
\usepackage{color}

\usepackage{etoolbox}

\AtBeginEnvironment{thebibliography}{\small}

\AtBeginEnvironment{thebibliography}{%
  \setlength{\itemsep}{0pt}%
  \setlength{\parskip}{0pt}%
}

\begin{document}

\title{Sensitivity Limits for $\phi$-OTDR}

\author{Juan M. Marin, Roman Ermakov, Huwei Wang, Francesco Da Ros, and Darko Zibar}

\affil{Department of Electrical and Photonics Engineering, Technical University of Denmark, Kongens Lyngby, Denmark}

\email{jmmmo@dtu.dk}

\begin{abstract}
We demonstrate a new lower bound on the sensitivity of $\phi$-OTDR by estimating phase using Kalman filtering. Numerical analysis shows a $\sim$35-dB reduction in the minimum detectable SNR, compared to when using conventional phase estimation. 
\end{abstract}

\section{Introduction}

Phase-sensitive optical time-domain reflectometry ($\phi$-OTDR) has become an established platform for distributed fiber-optic sensing. It effectively transforms every meter of an optical fiber into a sensing interface of mechanical and thermal phenomena over tens of kilometers~\cite{liu2025}. Consequently, this technology has been widely leveraged in applications such as perimeter surveillance and structural health monitoring \cite{martins2013}. In these applications, high sensitivity, as well as the prompt and precise location of abnormal events, are essential for enabling timely decisions and ensuring safety \cite{bao2021}. In practice, achieving these requirements is fundamentally limited by the accuracy of phase estimation from Rayleigh backscattering~\cite{gabai2016}.

However, the components of $\phi$-OTDR systems introduce noise into the phase estimation process. Certain noise sources, such as amplified spontaneous emission (ASE) from optical amplification and shot noise inherent to coherent detection, establish a minimum noise level that is always present in these systems, known as quantum noise \cite{martins2013, wang2025}. This, in turn, imposes a minimum induced phase fluctuation (uncertainty), and consequently a lower bound on $\phi$-OTDR sensitivity, commonly referred to as the quantum limit~\cite{wang2025, zibar2021}. 

Conventional phase estimation in $\phi$-OTDR consists of the calculation of the argument from Rayleigh backscattering using in-phase and quadrature (I/Q) detection~\cite{li2021}. Nonetheless, this approach does not account for noise and is suitable only when the embedded noise is negligible. This highlights the need for a statistically optimal phase estimator that minimizes uncertainty \cite{zibar2021}. Therefore, recursive estimators such as Kalman filters naturally fit this problem, and have already been explored in the literature~\cite{wang2024}. Their implementation in $\phi$-OTDR represents a significant sensitivity enhancement that, to the best of our knowledge, has not yet been quantified. 

Motivated by this gap, this work provides a numerical analysis to characterize the new lower bound in $\phi$-OTDR sensitivity that arises from using the Kalman filter as a statistically optimal phase estimator. In particular, we introduce a Sage--Husa adaptive Kalman filter (SHAKF) designed to dynamically estimate the noise statistics of any 
$\phi$-OTDR system. This overcomes the standard Kalman filter's reliance on guessed prior noise statistics, which can potentially introduce uncertainty~\cite{sage1969}.

\begin{figure}[t]
    \centering
    \includegraphics[width=\linewidth]{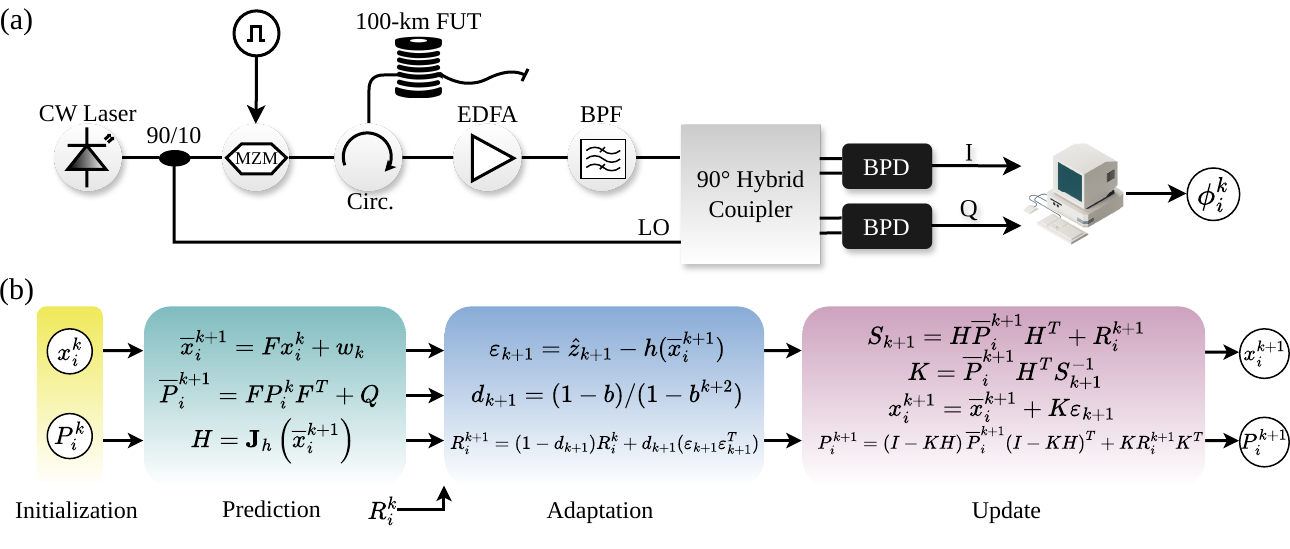}
    \caption{(a) Simulation setup for coherent $\phi$-OTDR. (b) Schematic of SHAKF phase estimation.}
    \label{fig1}
\end{figure}

\section{Data Generation}
\noindent To assess the impact of Kalman-based phase estimation in $\phi$-OTDR, the phase estimation performance of conventional I/Q demodulation is compared against SHAKF. For this purpose, I/Q data is generated by numerically simulating the coherent $\phi$-OTDR system illustrated in Fig.~\ref{fig1}(a). This system comprises a continuous-wave (CW) laser emitting 23~dBm of optical power at a wavelength of $\lambda_0 = 1550$~nm. This output is split using a 90/10 coupler. The first branch is modulated into a train of pulses by a Mach-Zehnder modulator (MZM) with a pulse duration of $\tau=10$~ns, and a repetition rate of $F_r=1$~kHz. The modulated signal is injected into a circulator (Circ.) and then into a fiber under test (FUT) with an attenuation coefficient $\alpha=0.2$~dB/km and a length of 100~km. The Rayleigh backscattering signal along the FUT is generated using the numerical model reported in \cite{Liokumovich2015}. 

This signal is then retrieved by the same circulator and amplified by an erbium-doped fiber amplifier (EDFA) with a gain of 22~dB and a minimum noise figure (NF) of 3~dB. After amplification, the Rayleigh backscattering is filtered by an optical bandpass filter (BPF) with a central wavelength matching $\lambda_0$, and a bandwidth of 0.1 nm. Moreover, the filtered signal is mixed with a local oscillator (LO) in a $90^{\text{o}}$ hybrid coupler, which represents the second branch of the 90/10 coupler. Each one of the outgoing signals (I and Q) is converted into the electrical domain via a balanced photodetector (BPD) with a bandwidth of 250~MHz, noise equivalent power (NEP) of 80~pW/$\sqrt{\text{Hz}}$, and a dark current of $I_d = 5$~nA. Ultimately, these signals are sampled by a digitizer with a sampling frequency of $F_s=500$~MSa/s. Using this model, a total of 250000 spatial points were sampled along 100 different FUTs over 500 frames.

\section{Phase Estimation}

The first approach consists in calculating the argument from Rayleigh backscattering by following $\phi_i^k = \arctan \left( \frac{Q_i^k}{I_i^k} \right)$, where the sub-indices $(i,k)$ represent the position along the fiber, and the frame number, respectively. On the other hand, SHAKF comprises an entire iterative algorithm, which is summarized in Fig. \ref{fig1}(b). This algorithm takes as input the system's noisy measurements $\hat{z}_k = \left[ I_i^k, Q_i^k \right]$. These values inherently convey information about variables of interest, such as the amplitude ($A_i^k$) and phase ($\phi_i^k$) of the Rayleigh backscattering constituting the hidden state $x_i^k = \left[ A_i^k, \phi_i^k\right]$.

The relationship between $x_i^k$ and $\hat{z}_k$ is modeled by $\hat{z}_k = h(\overline{x}_i^k) + r_k$, where $r_k$ is the system's noise, represented by a zero-mean Gaussian distribution with covariance $R_i^k$. The measurement function $h(x_i^k) = \left[ A_i^k \cos (\phi_i^k),\ A_i^k \sin (\phi_i^k) \right]$ allows $x_i^k$ to be estimated as a combination of incoming noisy measurements $\hat{z}_k$ and temporal predictions $\overline{x}_i^k$. At each iteration, SHAKF corrects its predictions using incoming measurements, quantifying the match between the predicted state and the new measurement via the innovation $\varepsilon_{k}$. With this in mind, the entire estimation process is divided into four steps: Initialization, Prediction, Adaptation, and Update. The first step initializes $x_i^k$ using values from preceding estimates, along with its associated covariance matrix $P_i^k$.

The second step predicts the evolution of $x_i^k$ and $P_i^k$ from one frame to another ($\overline{x}_i^{k+1}, \overline{P}_i^{k+1}$). The predictions are modeled by the state-transition matrix $F$, and $w_{k+1}$, which is a Gaussian, zero-mean process with an associated covariance matrix $Q$. In this scenario, $F$ is defined as the identity matrix. Therefore, the prediction model assumes that the temporal evolution of $x_i^k$ is a discrete-time Wiener process, which allows SHAKF to handle a wide range of events. In addition to this, the measurement function $h$ is linearized at the current estimate by calculating its Jacobian and evaluating it at $\overline{x}_i^{k+1}$.

Furthermore, the third step in the pipeline adapts SHAKF to the system's noise by iteratively adjusting the measurement noise covariance matrix $R_i^k$. The evolution of this tuning process from one frame to another $R_i^{k+1}$ is subjected to the amnestic factor $d_k$, and subsequently, to the designated forgetting factor $b$ (where $0<b<1$). A low $b$ makes SHAKF reliant on historical data trends while a high $b$ makes SHAKF reactive, giving more weight to recent measurements. The latter is best for prompt response to incoming events, hence we opted for $b=0.95$.

Finally, the last step provides a new estimate  ($x_i^{k+1}$, $P_i^{k+1}$), resulting from the combination of $\overline{x}_i^{k+1}$ and $\hat{z}_{k+1}$. The amount of trust deposited in each one is determined by the Kalman gain $K$. This value is calculated iteratively to correct SHAKF predictions based on $\overline{P}_i^{k+1}$ and $R_i^{k+1}$.

\section{Sensitivity Limit}

We assume there are no events or disturbances along the FUT, so the Rayleigh backscattering signal remains constant. The sensitivity of a $\phi$-OTDR system is defined as the minimum detectable phase change for a given signal-to-noise ratio (SNR) \cite{gabai2016}. To determine the sensitivity limit of each phase estimation method, we compute the phase uncertainty and SNR at each sampled point.

The phase uncertainty is calculated as the root mean square error (RMSE) according to $\sigma_\phi (i)  = \sqrt{\sum_{k=1}^N(\phi_i^k - \widetilde{\phi}_i^k)^2 / N}$, where $\widetilde{\phi}_i^k$ is the phase extracted from the same Rayleigh backscattering signal in absence of the system's noise, and $N$ the total number of frames.  The SNR is computed as $\text{SNR}(i) = |\overline{A_i}|^2/\sigma_A^2$, where $\overline{A_i}$ is the average Rayleigh backscattering intensity and $\sigma_A^2$ is its variance \cite{gabai2016, wang2025}. Under static conditions, this variance corresponds to the system's noise, which in our case is quantum noise. In the literature, the sensitivity of coherent $\phi$-OTDR in the quantum-noise regime is given by $\sigma_{\text{QL}} = \sqrt{1/(2\,\text{SNR})}$ \cite{wang2025}, or in dB, $\sigma_{\text{QL}} = -0.5\,\text{SNR} - 1.5~\text{dB}$.

\begin{figure}[t]
    \centering
    \includegraphics[width=\linewidth]{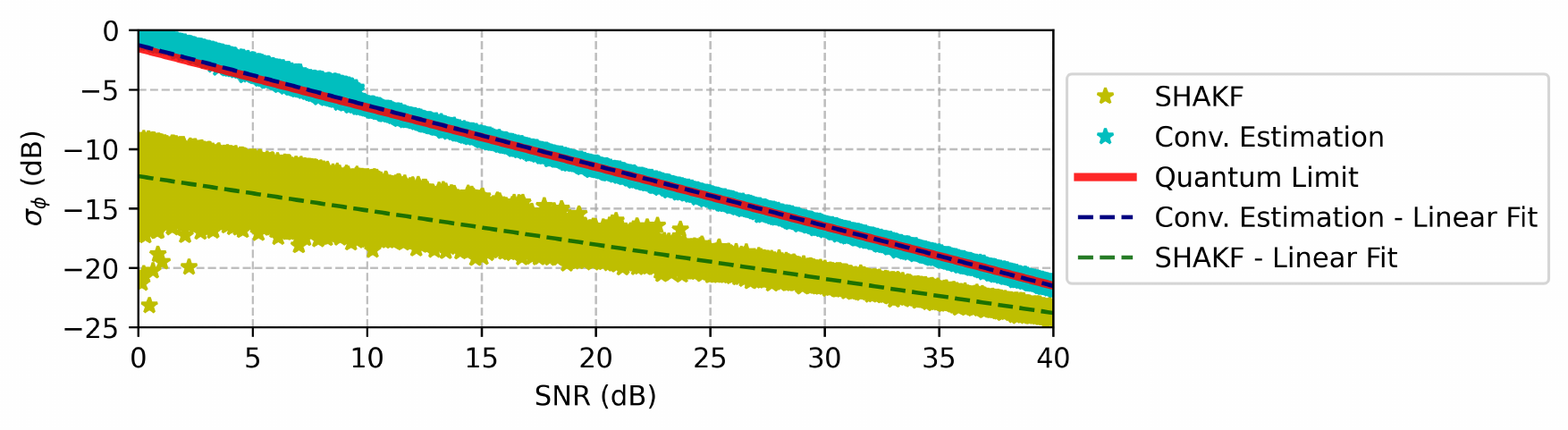}
    \caption{Phase uncertainty achieved by SHAKF and conventional phase estimation as a function of SNR, compared with the quantum limit.}
    \label{fig2}
\end{figure}

In Fig.~\ref{fig2}, we compare the sensitivity achieved by conventional phase estimation against SHAKF. Conventional estimation closely follows its theoretically derived quantum limit, with a linear fit of $\sigma_{\text{I/Q}} = -0.5\,\text{SNR} - 1.26~\text{dB}$. In contrast, SHAKF achieves substantially lower phase uncertainty, with a fitted curve of $\sigma_{\text{SHAKF}} = -0.29\,\text{SNR} - 12.28~\text{dB}$, showing both a reduced slope and lower intercept. 

For this analysis, it is beneficial to establish a maximum tolerable phase uncertainty. Defining this threshold enables the inference of the minimum SNR at which phase can be estimated reliably ($\text{SNR}_{th}$). Accordingly, we define a phase uncertainty threshold of $\sigma_{\text{th}} = -3$~dB. Using this limit, $\text{SNR}_{th}$ is extrapolated from the $\sigma_{\text{QL}}$, $\sigma_{\text{I/Q}}$, and $\sigma_{\text{SHAKF}}$ curves as a performance metric. Following this approach, the theoretical and calculated $\text{SNR}_{th}$ values for conventional phase estimation are 3~dB and 3.5~dB, respectively. In contrast, the $\text{SNR}_{th}$ achieved by SHAKF drops drastically to $-32$~dB. This represents a $\sim$35~dB improvement in sensitivity, defining a new lower bound for phase estimation.

\section{Conclusions}
We demonstrated a new lower bound in 
$\phi$-OTDR sensitivity using Kalman-based phase estimation instead of the conventional method. Results show phase-estimation uncertainty below -3 dB for points on a fiber with an SNR as low as -32 dB. This underscores the need to view phase estimation as an uncertainty-minimization problem, which can be addressed with a statistically optimal estimator such as the Kalman filter. These results reveal the potential to reach longer monitoring distances and open new doors for distributed sensing applications.

\printbibliography

@article{bao2021,
  title={Recent advancements in Rayleigh scattering-based distributed fiber sensors},
  author={Bao, Xiaoyi and Wang, Yuan},
  journal={Advanced devices \& instrumentation},
  year={2021},
  publisher={AAAS}
}

@article{gabai2016,
  title={On the sensitivity of distributed acoustic sensing},
  author={Gabai, Haniel and Eyal, Avishay},
  journal={Optics letters},
  volume={41},
  number={24},
  pages={5648--5651},
  year={2016},
  publisher={Optical Society of America}
}

@inproceedings{wang2025,
  title={Quantum Noise Limited Temperature-Change Estimation for $\phi$-OTDR Employing Coherent Detection},
  author={Wang, Huwei and Ermakov, Roman and Da Ros, Francesco and Zibar, Darko},
  booktitle={2025 European Conference on Optical Communications (ECOC)},
  pages={1--4},
  year={2025},
  organization={IEEE}
}

@article{martins2013,
  title={Coherent noise reduction in high visibility phase-sensitive optical time domain reflectometer for distributed sensing of ultrasonic waves},
  author={Martins, Hugo F and Martin-Lopez, Sonia and Corredera, Pedro and Filograno, Massimo L and Fraz{\~a}o, Orlando and Gonz{\'a}lez-Herr{\'a}ez, Miguel},
  journal={Journal of Lightwave Technology},
  volume={31},
  number={23},
  pages={3631--3637},
  year={2013},
  publisher={IEEE}
}

@article{zibar2021,
  title={Approaching optimum phase measurement in the presence of amplifier noise},
  author={Zibar, Darko and Pedersen, Jens E and Varming, Poul and Brajato, Giovanni and Da Ros, Francesco},
  journal={Optica},
  volume={8},
  number={10},
  pages={1262--1267},
  year={2021},
  publisher={Optical Society of America}
}

@article{li2021,
  title={Phase demodulation methods for optical fiber vibration sensing system: A review},
  author={Li, Yan and Wang, Yu and Xiao, Lin and Bai, Qing and Liu, Xin and Gao, Yan and Zhang, Hongjuan and Jin, Baoquan},
  journal={IEEE Sensors Journal},
  volume={22},
  number={3},
  pages={1842--1866},
  year={2021},
  publisher={IEEE}
}

@article{wang2024,
  title={Interference fading suppression with fault-tolerant Kalman filter in phase-sensitive OTDR},
  author={Wang, Yu and He, Chunchen and Du, Waner and Hu, Huirong and Bai, Qing and Liu, Xin and Jin, Baoquan},
  journal={Isa Transactions},
  volume={150},
  pages={298--310},
  year={2024},
  publisher={Elsevier}
}

@inproceedings{sage1969,
  title={Adaptive filtering with unknown prior statistics},
  author={Sage, Andrew P and Husa, Gary W},
  booktitle={Joint Automatic Control Conference},
  number={7},
  pages={760--769},
  year={1969}
}

@article{Liokumovich2015,
  author={Liokumovich, Leonid B. and Ushakov, Nikolai A. and Kotov, Oleg I. and Bisyarin, Mikhail A. and Hartog, Arthur H.},
  journal={Journal of Lightwave Technology}, 
  title={Fundamentals of Optical Fiber Sensing Schemes Based on Coherent Optical Time Domain Reflectometry: Signal Model Under Static Fiber Conditions}, 
  year={2015},
  volume={33},
  number={17},
  pages={3660-3671}
}

@article{liu2025,
  title={Advances in phase-sensitive optical time-domain reflectometry},
  author={Liu, Shuaiqi and Yu, Feihong and Hong, Rui and Xu, Weijie and Shao, Liyang and Wang, Feng},
  journal={Opto-Electronic Advances},
  volume={5},
  number={3},
  pages={200078--1},
  year={2025}
}

\end{document}